\documentclass[12pt,twoside]{article}   %For printing on two sides of page
\usepackage[super,sort,comma]{natbib}

\usepackage{fancyhdr}		%Gives headers and footers defined below
\usepackage[section]{placeins}   %

\usepackage[title]{appendix}

\usepackage{graphicx}
\usepackage{multirow}

\makeatletter \renewcommand\@biblabel[1]{$^{#1}$} \makeatother
\newcommand{\cen}[1]{\begin{center} #1 \end{center}}

\usepackage{isotope}
\usepackage{xspace}

\usepackage{hyperref}
\hypersetup{ colorlinks,
	citecolor=blue,
	filecolor=blue,
	linkcolor=blue,
	urlcolor=blue
}

\usepackage{xcolor}
\definecolor{gray}{rgb}{0.6,0.6,0.6}
\definecolor{red}{rgb}{0.85,0,0}
\definecolor{green}{rgb}{0,0.85,0}
\definecolor{blue}{rgb}{0,0,0.85}
\definecolor{beige}{rgb}{0.92,0.87,0.78}
\definecolor{amaranth}{rgb}{0.9, 0.17, 0.31}
\definecolor{magenta}{rgb}{0.6, 0.4, 0.8}
\usepackage[all]{hypcap}    %causes link to figures to go to figure, not caption
\usepackage{amsmath}
\usepackage{amssymb}
\usepackage[normalem]{ulem}
\usepackage{upgreek}
\usepackage{comment}

\begin{document}

%\input{newcommands.tex}
% new commands 
\newcommand{\Hidex}{\textrm{Hidex}\xspace}
\newcommand{\leakage}{$\ensuremath{P}_{leak}(Pb)\xspace$}
\newcommand{\lfrac}{$\ensuremath{f}_{leak}(Pb)\xspace$}

%colors
\newcommand{\impact}{\textit{\textbf{b}}\xspace}
\newcommand{\red}{\color{red}}
\newcommand{\blue}{\color{blue}}	 
\newcommand{\green}{\color{ForestGreen}} 
\newcommand{\magenta}{\color{magenta}}
\newcommand{\orange}{\color{orange}}

%math symbols
\newcommand{\lradon}{$\ensuremath{L}_{Rn} \xspace$}
\newcommand{\lradoneff}{$\ensuremath{L}_{Rn}^{eff}$}
\newcommand{\llead}{$\ensuremath{L}_{Pb} \xspace$}
\newcommand{\lbis}{$\ensuremath{L}_{Bi} \xspace$}
\newcommand{\ldiff}{$\ensuremath{L}_{eff}$}
\newcommand*{\avgldiff}{$\langle \ensuremath{L}_{eff}\rangle \xspace$}
\newcommand*{\dart}{Alpha-DaRT }
\newcommand{\radium}{$\isotope[224]{Ra}$}
\newcommand{\radon}{$\isotope[220]{Rn}$}
\newcommand{\lead}{$\isotope[212]{Pb}$}
\newcommand{\thorium}{$\isotope[228]{Th}$}
\newcommand{\polonium}{$\isotope[216]{Po}$}
\newcommand{\poloniumm}{$\isotope[212]{Po}$}
\newcommand{\bismuth}{$\isotope[212]{Bi}$}
\newcommand{\thelurium}{$\isotope[208]{Tl}$}
\newcommand{\cesium}{$\isotope[137]{Cs}$}
\newcommand{\cobalt}{$\isotope[60]{Co}$}

\cen{\sf {\Large {\bfseries Diffusing Alpha-emitters Radiation Therapy:
In vivo Measurements of Effective Diffusion and Clearance Rates Across Multiple Tumor Types} \\  
		\vspace*{10mm}
		 Mirta Duman\v{c}i\'{c}$^{1,*,\dag}$, Guy Heger$^{1,*}$, Ishai Luz$^{2}$, Maayan Vatarescu$^{2}$, Noam Weizman$^{1,3}$, Lior Epstein$^{1,4}$, Tomer Cooks$^{2,\ddag}$, Lior Arazi$^{1,\ddag}$} \\
  \vspace{5mm}
	$^1$Unit of Nuclear Engineering, Faculty of Engineering Sciences, Ben-Gurion University of the Negev, P.O.B. 653 Be'er-Sheva 8410501, Israel \\
     $^2$ The Shraga Segal Department of Microbiology, Immunology, and Genetics, Ben-Gurion University of the Negev, Be'er Sheva, Israel \\
     $^3$ Radiation Therapy Unit, Oncology Department, Hadassah Medical Center, Jerusalem 91120, Israel \\
     $^4$ Radiation Protection Department, Soreq Nuclear Research Center, Yavne, Israel \\
     $^*$ These authors contributed equally \\
     $^\dag$ Current affiliation: Medical Physics Unit, Gerald Bronfman Department of Oncology, Faculty of Medicine and Health Sciences, McGill University, Montreal, Quebec, Canada
      \vspace{5mm}\\
        Version typeset \today\\
}

\pagenumbering{arabic}
\setcounter{page}{1}
\pagestyle{plain}
$^\ddag$ Authors to whom correspondence should be addressed. Lior Arazi: larazi@bgu.ac.il, Tomer Cooks: cooks@bgu.ac.il\\

\begin{abstract}
\noindent 
    {\bf Background:} Diffusing alpha-emitters radiation therapy (``Alpha-DaRT'') is a new modality that uses alpha particles to treat solid tumors. Alpha-DaRT employs interstitial sources loaded with low activities of $^{224}$Ra, which release a chain of short-lived alpha-emitters diffusing over a few millimeters around each source. Alpha-DaRT dosimetry is described, to first order, by a framework called the ``diffusion-leakage'' (DL) model. \\
    {\bf Purpose:} The aim of this work is to estimate the tumor-specific parameters of the DL model from \textit{in vivo} studies on multiple histological cancer types. \\
    {\bf Methods:} Autoradiography studies with phosphor imaging were conducted on 113 mice-borne tumors from 10 cancer cell lines. An observable, referred to as the ``effective diffusion length'' \ldiff, was extracted from images of histological slices obtained using phosphor screens. The tumor and Alpha-DaRT source activities were measured after excision with a gamma counter to estimate the probability of \lead~clearance from the tumor by the blood, \leakage. \\
    {\bf Results:} The measured values of \ldiff~are in the range of 0.2-0.7~mm across different tumor types and sizes. \leakage~ is between 10–90\% for all measured tumors and it generally decreases in magnitude and spread for larger tumors. \\
    {\bf Conclusions:} The measured values of \ldiff~and \leakage~and associated dose calculations indicate that hexagonal Alpha-DaRT source lattices of $\sim4$~mm spacing with $\mu$Ci-scale $^{224}$Ra activities can lead to effective coverage of the tumor volume with therapeutic dose levels, with considerable margin to compensate for local variations in diffusion and leakage. \\

\end{abstract}

%
% Uncomment for keywords
\vspace{2pc}
\newpage

\section{Introduction} \label{section:intro}

Alpha particles are of great interest for radiotherapy due to their high relative biological effectiveness (RBE), their efficacy against hypoxic cells, and the potential for sparing healthy tissue ~\cite{mird:2010}. 
Multiple preclinical and clinical studies are underway in targeted alpha-therapy (TAT)~\cite{tafreshi:2019, jama:2018, revtat:2018, revtat2:2018, revtat3:2019} with $^{223}$RaCl$_2$ already FDA-approved for clinical use against bone metastases in castration-resistant prostate cancer~\cite{xofigo:2014, pacilio:2015, chittenden:2016, vera:2021}. Despite their benefits, the short range of alpha particles has largely hindered their clinical use against solid tumors. Only recently, clinical trials using PSMA-TAT~\cite{ac:2021} have begun showing successful results against macroscopic metastases.

A different approach for treating solid tumors with alpha radiation is \textit{diffusing alpha-emitters radiation therapy} (``Alpha-DaRT'')~\cite{arazi:2007}. Alpha-DaRT is based on intratumoral insertion of sources loaded with a few $\mu$Ci of radium-224 (\isotope[224]{Ra}, $t_{1/2}=3.63$\;d), that continuously release from their surface a chain of short-lived alpha- and beta--emitting atoms (\isotope[220]{Rn}, \isotope[216]{Po}, \isotope[212]{Pb}, \isotope[212]{Bi}, \isotope[212]{Po} and \isotope[208]{Tl}). 
Those daughter atoms spread throughout the tumor by diffusion and (potentially) vascular and interstitial convection, creating a therapeutic region of several mm in diameter around each source through their alpha decays and so overcoming the short-range limitation of alpha particles~\cite{arazi:2007}.  Alpha-DaRT has been studied \textit{in vitro} and \textit{in vivo} as a stand-alone treatment~\cite{arazi:2007, cooks:2008, arazi:2010, cooks:2009, lazarov:2012, cooks:2012} and in combination with chemotherapy~\cite{cooksCancer:2009, drori:2012, milrot:2013, reitkopf:2015} and immunotherapy~\cite{keisari:2013, confino:2015, confino:2016, vered:2019, vered:2020, keisari:2020, keisari:2021, vered:2022}. However, no systematic study has been conducted so far to compare the physics properties of Alpha-DaRT treatment across various histological cancer types.

In the first clinical trial on locally advanced and recurrent squamous cell carcinoma (SCC) of the skin and head and neck~\cite{popovtzer}, all treated tumors showed a positive response (30--100\% shrinkage in their longest dimension) and nearly 80\% displayed complete response. The Alpha-DaRT source $^{224}$Ra activity of $2~\mu$Ci/cm and nominal spacing of 5~mm between the sources were selected based on dose calculations performed by an approximate theoretical approach---the ``diffusion-leakage'' (DL) model \cite{arazi:2020, latticepart1:2023, latticepart2:2023}. The calculations relied on preliminary estimates for the DL model parameters for SCC, derived from autoradiography experiments on such tumors in mice \cite{arazi:2020} and aimed to cover the tumor volume with a minimal alpha dose of 10~Gy. A following pilot study in the US, on patients with recurrent or unresectable skin cancer (SCC and basal cell carcinoma), employed sources of $3~\mu$Ci/cm and nominal spacing of 4~mm, reporting 100\% complete response at 12 weeks following treatment (with 90\% confirmed by CT)~\cite{DAndrea2023}. A more recent publication, summarizing a pooled analysis of 81 treated head and neck or skin tumors from four clinical trials with a median follow-up of 14 months, reported a complete response in 89\% of the treated lesions, a two-year local recurrence-free survival of 77\%, and no grade 2 or higher late toxicities \cite{Popovtzer2024}. With more clinical trials underway (e.g., for pancreatic cancer \cite{Miller2024}), there is a need to estimate the DL model parameters in additional tumor types to recommend a plausible starting point for treatment planning in terms of source activity and spacing.
Different tumor models offer various conditions of the tumor microenvironment and tissue perfusion (e.g., blood flow, blood volume, mean transit time) that can be expected to affect the relevant DL model parameters~\cite{baker:2011}. Cell-type specific DNA repair rates and immunogenic factors can all contribute to a variable rate of cell death in each tumor model and hence cause different levels of necrosis in the tumor.

In this study, we report on results from a systematic campaign of \textit{in vivo} experiments with Alpha-DaRT treatment in 113 mice tumors across 10 different cell lines, including breast cancer, prostate cancer, pancreatic cancer, and colorectal cancer cell lines. The experiments were done after 4--5 days of treatment when all daughter atoms are in secular equilibrium with \isotope[224]{Ra}, mirroring the situation in real-life clinical trials. A recently published study\cite{Heger2024} focused on measuring diffusion properties for short source dwelling times, where only \isotope[220]{Rn} and \isotope[216]Po are in secular equilibrium with \isotope[224]{Ra}. The two complementary studies combined provide an experimental basis---albeit limited to mice---for suggesting a reasonable starting point for treatment planning in future Alpha-DaRT clinical trials for different tumor indications.   \par

\section{Methods}
\subsection{The Diffusion-Leakage model in equilibrium} \label{section:DLModel}
In a previously published work~\cite{arazi:2020}, it was shown that the two most important parameters governing the spread of activity inside the tumor are the diffusion lengths of \radon~and \lead, \lradon~and \llead. In a recently published first estimate of \lradon~in 24 mice-borne subcutaneous tumors with five different cell lines~\cite{Heger2024}, the source dwelling time inside the tumor was kept to 30~min, thus preventing the buildup of \lead. The measured values were in the range $L_{Rn}=0.25-0.55$~mm, with similar results obtained \textit{in-vivo} ($L_{Rn}=0.40\pm0.08$~mm, $n=14$) and \textit{ex-vivo} ($L_{Rn}=0.39\pm0.07$~mm, $n=10$).\footnote{In the \textit{ex-vivo} experiments, performed to measure daughter atom migration in the absence of blood and interstitial flows, sources were inserted into the tumor immediately after their removal, with the tumors kept in PBS inside an incubator at $37^\circ$C for 30 minutes.} To estimate the contribution of \llead, however, the source dwelling time should be several days, such that \lead~reaches secular equilibrium with \radium. Under such conditions, the Alpha-DaRT diffusion process affecting the overall treatment volume is a combination of two relevant length scales governing the radon and lead diffusion in tissue. On the relevant timescale of several days ($t \gg \tau_{Rn}, \tau_{Pb}$), the specific activities of \radon~and \lead~for a point source take on the asymptotic form:

\begin{equation}
    \lambda_{Rn}n^{asy}_{Rn}(r, t) =\lambda_{Rn} A_{Rn} \frac{e^{-r/L_{Rn}}}{r} e^{-\lambda_{Ra}t},
     \label{eq:radon}
\end{equation}

\begin{equation}
    \lambda_{Pb}n^{asy}_{Pb}(r, t) =\lambda_{Pb} \left( A_{Pb} \frac{e^{-r/L_{Rn}}}{r} + B_{Pb} \frac{e^{-r/L_{Pb}}}{r} \right) e^{-\lambda_{Ra}t}, \label{eq:lead}
\end{equation}
where $n_{Rn}, n_{Pb}$ are the number densities of \radon~and \lead, $\lambda_{Ra}, \lambda_{Rn}, \lambda_{Pb}$ are the decay rate constants of \radium, \radon~and \lead, and $A_{Rn}$, $A_{Pb}$ and $B_{Pb}$ are constants determined by the source and tissue properties. The radial activity distributions around the source are governed by \lradon~and \llead, given by:

\begin{equation}
    L_{Rn}=\sqrt{\frac{D_{Rn}}{\lambda_{Rn}-\lambda_{Ra}}} \approx \sqrt{D_{Rn}\tau_{Rn}} ,
    \label{eq:ldiff_radon}
\end{equation}

\begin{equation}
    L_{Pb}=\sqrt{\frac{D_{Pb}}{\lambda_{Pb}+\alpha_{Pb}-\lambda_{Ra}}} \approx \sqrt{D_{Pb}\tau_{Pb}^{eff}},
    \label{eq:ldiff_lead}
\end{equation}
where $D_{Rn}$ and $D_{Pb}$ are the effective diffusion coefficients of \radon~and \lead, $\tau_{Rn}~=~1/\lambda_{Rn}$ is the mean lifetime of \radon, $\alpha_{Pb}$ is the clearance rate constant of \lead~through the blood and $\tau_{Pb}^{eff}=\frac{1}{\lambda_{Pb}+\alpha_{Pb}}$ is the effective mean lifetime of \lead, accounting for both radioactive decay and clearance. The total asymptotic alpha dose is also governed by the exponential terms appearing in Equation~\ref{eq:lead}~as previously shown\cite{arazi:2020}. For long cylindrical sources, these exponents are replaced by the modified Bessel function of the second kind $K_{0}(r/L)$~\cite{latticepart1:2023}.\par 

\begin{figure}
\includegraphics[width=0.9\textwidth]{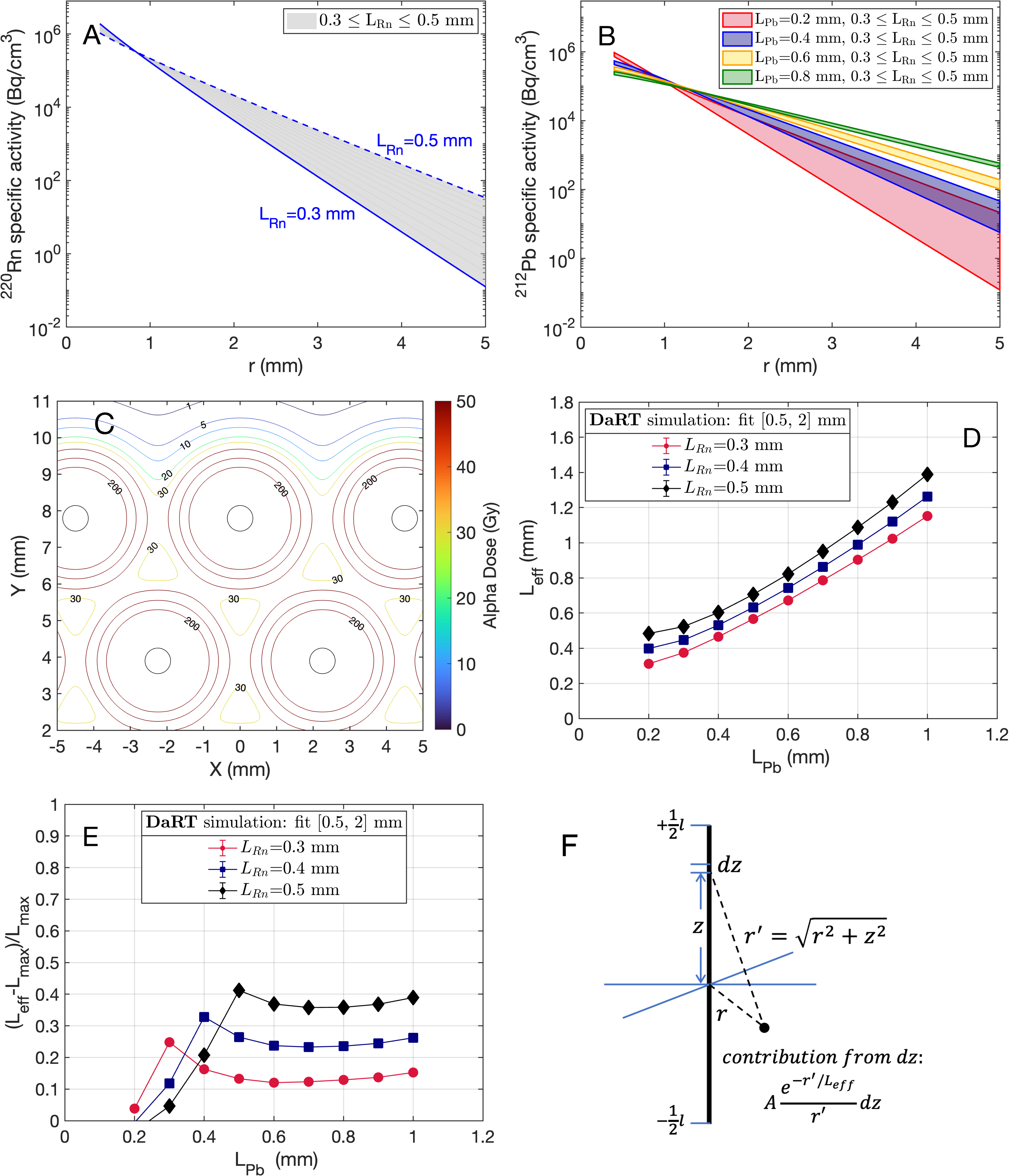}
\caption{Specific activities of \radon~(A) and \lead~(B) for $0.3\leq $\lradon$\leq 0.5$~mm, and in panel B, $L_{Pb}=0.2, 0.4, 0.6, 0.8$~mm. (C) Asymptotic alpha dose in a hexagonal lattice with 4.5~mm spacing for $L_{Rn}=0.4$~mm and $L_{Pb}=0.2$~mm. The source \radium~ activity in (A)-(C) is $3~\mu$Ci/cm, with $P_{des}(Rn)=0.45$, $P_{des}^{eff}(Pb)=0.55$ and $P_{leak}(Pb)=0.5$. (D) Estimated \ldiff~ and (E) the relative difference with respect to the dominant diffusion length vs. the simulated \llead~for \lradon=0.3, 0.4 and 0.5~mm. (F) Illustration of the infinitesimal contribution to the integral in the line source approximation in Equation~\ref{eq:fitmodel}.}

\label{fig:model_4dy}
\end{figure}

Figure~\ref{fig:model_4dy}A and B show the radial specific activity distributions of \radon~and \lead~four days after source insertion, as obtained by numerically solving the full DL equations for an infinite 0.7~mm-diameter cylindrical source (it was shown that in the source midplane, the infinite- and finite-cylinder solutions converge to $<1\%$). A full discussion of the solution can be found in Refs.~\cite{latticepart1:2023, latticepart2:2023}. The initial source \radium~activity per unit length was set to $3~\mu$Ci/cm. The \lead~leakage probability $P_{leak}=\alpha_{Pb}/(\lambda_{Pb}+\alpha_{Pb})$ (i.e., the probability that a \lead~atom released from the source decays outside of the tumor) was set to 0.5, as a moderate case of blood clearance. The respective desorption probabilities of \radon~and \lead~(i.e. the probability that a decay of \radium~on the source leads to their release into the tumor) were set as $P_{des}(Rn)=0.45$ and $P_{des}^{eff}(Pb)=0.55$, which are typical properties of the sources provided by the manufacturer, Alpha Tau Medical Ltd. The \bismuth~diffusion length and clearance rate constant were set as $L_{Bi}=0.1L_{Pb}$ and $\alpha_{Bi}=0$, following arguments discussed in previous publications~\cite{arazi:2020}. The \radon~specific activity around the DaRT source is shown for $0.3\leq L_{Rn}\leq0.5$~mm, consistent with the range recently reported by Heger \textit{et al}~\cite{Heger2024}. A spread of about one order of magnitude in specific activity is observed at 3~mm from the source within the measured values of $L_{Rn}$ (Figure~\ref{fig:model_4dy}A). The \lead~activity distribution is shown separately for $L_{Pb}=0.2, 0.4, 0.6$ and $0.8$~mm while varying \lradon~within the same range (Figure~\ref{fig:model_4dy}B). One sees a different magnitude of the activity spread as \llead~becomes the dominant scaling parameter: for \llead=0.2~mm the effect of varying \lradon~covers a spread of $\gtrsim1$ order of magnitude at distances 3--4~mm from the source, while for \llead=0.6--0.8~mm, the distribution of \lead~is almost completely independent of \lradon. \\ 

In order to illustrate the therapeutic effect for a particular set of DL parameters, Figure~\ref{fig:model_4dy}C shows the asymptotic alpha particle dose calculated for a hexagonal lattice of 3~$\mu$Ci/cm \radium~\dart sources at 4.5~mm spacing with $L_{Rn}=0.4$~mm and $L_{Pb}=0.2~$mm (other model parameters are as described above). The dose is calculated by accounting for contributions from all alpha decays in the radioactive chain. A therapeutic nominal alpha dose above 20~Gy~\cite{popovtzer} is administered to the tumor up to 2~mm away from the outermost sources. Beyond $\sim$3~mm away from the lattice, lower dose levels ($<$1~Gy) are beneficial for sparing healthy tissue.

\subsection{Effective diffusion length}

After 4--5 days of \dart treatment, the spread of activity inside the tumor is governed by both \lradon~and \llead, as shown in Equation~\ref{eq:lead} and illustrated in Figure~\ref{fig:model_4dy}B. However, when considering the radial distribution of activity around the source, the determining factor is the dominant diffusion length (either $L_{Rn}$ or $L_{Pb}$). In practice, this is embodied by the use of an \textit{effective} diffusion length, \ldiff, which can correspond to either the dominant diffusion length (if there is one) or be comprised of contributions from both if they are of comparable value. Note that \ldiff~is always the larger of the two diffusion lengths if the spatial profile is fitted at a sufficiently large distance from the source; deviations from this rule are observed if the fit is done at short distances, which is sometimes dictated by the size of the tumor or the quality of the tissue section. In addition, if diffusion is the main, but not only, process occurring inside the tumor, \ldiff~can also be affected by the contribution of other processes that are not currently included in the DL model (e.g. anisotropic vascular convective effects). To avoid potential bias by such effects, which generally appear farther away from the source and tend to increase \ldiff, the fit region is conservatively limited to small radial distances from the source (0.5--2~mm). Under this approximation, the asymptotic distribution of \lead~activity for a point source is given by:

\begin{equation}
    \lambda_{Pb}n^{asy}_{Pb}(r, t) \approx \lambda_{Pb} \, C_{Pb} \frac{e^{-r/L_{eff}}}{r} e^{-\lambda_{Ra}t},
\end{equation}
where the spatial activity profile is determined by one parameter \ldiff. This parameter can now be estimated by measuring the spatial distribution of $^{212}$Pb or moreover, the $^{212}$Bi progeny that appears to be in local secular equilibrium~\cite{arazi:2020}. This allows us to utilize the autoradiography approach based on recording the alpha particles emitted by \bismuth~and \poloniumm.

The accuracy of the one-parameter \ldiff~approach was assessed by simulation within the DL framework in the following manner. A full numerical calculation was performed for an infinite cylindrical source, varying the diffusion lengths over the ranges $0.3\leq L_{Rn}\leq0.5$~mm and $0.2\leq L_{Pb}\leq1.0$~mm, with the other model parameters as above. As done in Ref. \cite{Heger2024}, to obtain \ldiff~from the simulated data, an approximate analytical expression was used for fitting the solution of the DL equations, with the source considered as a finite line source consisting of point-like segments (this is justified since the slope of the logarithmic fall-off of the activity does not depend on the source diameter \cite{latticepart1:2023}). The specific activity of \bismuth~in the source midplane was extracted in the simulation at $t=4$~days and fitted assuming the line-source model: 

\begin{equation}
    \lambda_{Bi}n^{asy}_{Bi}(r) = A \int_{-l/2}^{l/2} \frac{e^{-\sqrt{r^2+z^2}/L_{eff}}}{\sqrt{r^2+z^2}}dz,
    \label{eq:fitmodel}
\end{equation}
where the integral is along the source's 6.5~mm length (as used in \textit{in vivo} experiments), summing contributions from point-like elements $dz$, with and $A$ and \ldiff~as free parameters (see Figure~\ref{fig:model_4dy}F). Four different fit regions were defined starting from 0.5~mm and ending with r=1~mm, r=1.5~mm, r=2~mm and r=2.5~mm. 
Figure~\ref{fig:model_4dy}D shows the fitted \ldiff~for the region $r=[0.5, 2]$~mm as a function of \llead~for $L_{Rn}=0.3, 0.4$ and $0.5$~mm, with Figure~\ref{fig:model_4dy}E showing the relative deviation of \ldiff~from the dominant simulated diffusion length. For $L_{Rn}=0.4$~mm and $L_{Pb}\gtrsim0.4$~mm, the fitted $L_{eff}$ can overestimate the dominant diffusion length by up to $\sim30\%$. For cases where \lead~ diffusion is sub-dominant ($L_{Pb}<L_{Rn}$), \ldiff~can overestimate $L_{Rn}$ on the level of $\sim10-20\%$. For $L_{Pb}\approx0.2$~mm, \ldiff~is consistent with \lradon~ to a few \%.

\subsection{Experimental Materials and Procedures} 
\label{sec:methods}

%\paragraph{Cells, tumors and sources.}
%\label{sec:cell_lines}

The details of the methodology and materials for growing cell cultures and tumor inoculation are given in Appendix~\ref{app:methods}. The cell lines used were: SQ2 (murine) for SCC (including a reanalysis of data obtained in previous studies at Tel Aviv University (TAU)), MDA-MB-231 (human) and 4T1 (murine) for triple-negative breast cancer (TNBC), PANC1 (human) and PANC2 (murine) for pancreatic duct adenocarcinoma (PDAC), MC38 (murine) and HCT116 (human) for colon carcinoma, and PC3, C4-2 and LNCaP (all human) for prostate carcinoma. All tumors were grown subcutaneously (except for five orthotopic 4T1 tumors) and treated with a single Alpha-DaRT source, inserted by a uniquely designed preclinical applicator. Tumors were grown across a range of sizes in order to inspect the dependence of the diffusion parameters on the tumor mass, under the expectation that larger tumors are characterized by a larger necrotic volume. In all recent experiments, the sources---made of 316LVM stainless steel---were loaded with 3~$\mu$Ci \isotope[224]{Ra}, and were 0.7 mm in diameter and 6.5 mm in length~\cite{veredgbm:2022}. In the previous studies at TAU, the sources carried $0.5-3.3~\mu$Ci \isotope[224]{Ra}, and were 0.3 mm in diameter and 5 mm long. The manufacturer guarantees that source activity and desorption probability are within 10\% of their nominal values. In a subgroup of these experiments (referred to as ``TAU point source'') the activity was limited to $\sim$1 mm from the source tip. \\
The autoradiography measurement procedure was similar to the one described in a previous publication~\cite{arazi:2007}. Histological sections (10-$\mu$m thick), cut approximately perpendicular to the source axis and laid on glass slides were placed on a  $20\times40$~cm$^2$ tritium-sensitive phosphor-imaging plate (BAS-IP TR2040S, Fujifilm, Japan) over a 12~$\mu$m Mylar foil to prevent direct contact with the plate, and left for a one-hour-long exposure. The plate was then scanned by an image reader (FLA-9000, Fujifilm, Japan) obtaining a gray-level (GL) output image with 100~$\mu$m pixels. For the additional SQ2 dataset, the scanner used was FLA-2000. The conversion from GL to photo-stimulated luminescence (PSL) and then into activity is described in Appendix~\ref{app:calibration}. As noted above, since \lead~($t_{1/2}=10.62$~h) is a beta emitter, and \radon~will have completely decayed by the time the slides are placed on the phosphor plate ($t_{1/2} = 55.6~s$), what is being imaged during the exposure are in fact the alpha decays of \bismuth~($t_{1/2}=60.55$ min) and \poloniumm~($t_{1/2}=0.30~\mu$s), which occur essentially at the same place where $^{212}$Pb decays. 

On average 5--7 glass slides were processed from the center of each tumor. Each slide had 1--3 consecutive histological slices. A section was excluded from the analysis in case of severe tissue damage giving rise to a distorted activity distribution. An additional requirement was that the total PSL count of the section is $>$800 (estimated as the noise level from the tail of the total PSL distribution across different tumors). Finally, tumors that had at least two histological sections passing all the requirements were considered for the final selection. To account for possible machine-dependent differences in image production, raw PSL images recorded by the FLA-9000 phosphor-imaging system were cross-calibrated with the iQID camera~\cite{miller2014} using the exact same tumor sections. The latter device does not require image deconvolution due to the event-by-event imaging of alpha decays (with an accuracy of $\sim20~\mu$m for each alpha-particle hit), and therefore allows for direct \ldiff~estimation from the recorded image. However, due to the longer exposure times (12--24 h) and limited active area (5-cm diameter) of the iQID system employed here, its throughput is much lower compared to the $40\times20$~cm$^2$ phosphor imaging plate. A detailed explanation of the cross-calibration can be found in Appendix ~\ref{app:calibration}. An overall correction factor of $1.09\pm0.11$ was applied per tumor to \ldiff ~measured from phosphor imaging. \par

\subsection{Estimating \ldiff~from data} 
To estimate \ldiff~from the spatial PSL distribution in the tumor section, the source position was determined either manually when the source left a recognizable central opening (``hole'') in the tumor tissue, or by calculating the center of gravity of all pixels in the ``hottest'' region of the image weighted by their PSL level. The measured specific activity profile, averaged over concentric rings centered on the estimated source position, was fitted by the model with \ldiff~as a free parameter according to the procedure explained in Section~\ref{section:DLModel} and using Equation~\ref{eq:fitmodel}, by employing the `lsqnonlin' Matlab function with a linear-log regression model (that uses the Levenberg-Marquardt algorithm~\cite{10.1007/BFb0067700}). A comparison between the two methods for estimating the source position on a few representative cases showed that their respective estimates for \ldiff~are consistent to within $\lesssim10\%$.

\begin{figure}[ht!]
\centering
\includegraphics[width=0.7\textwidth]{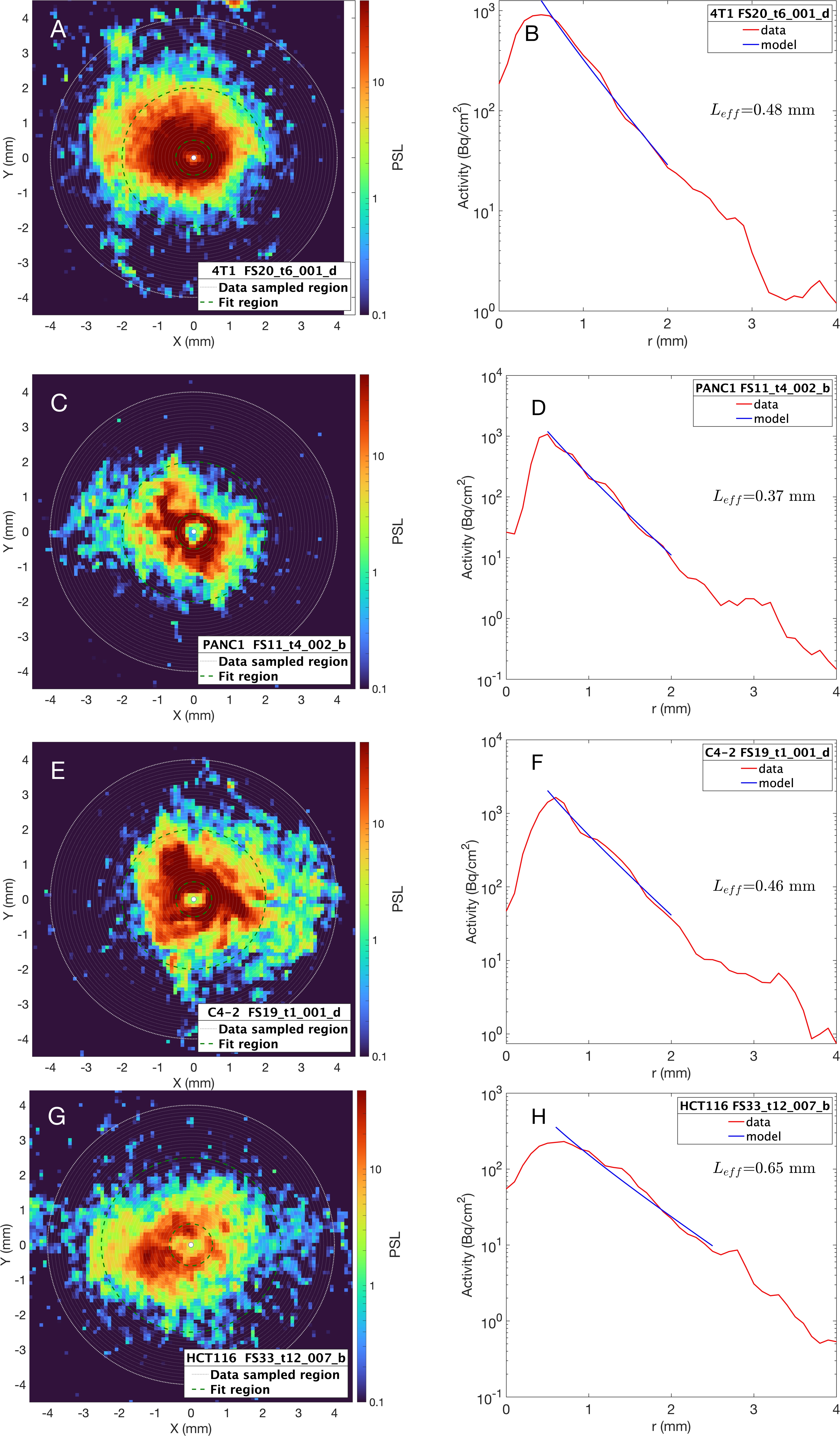}
\caption{Examples of PSL images from histological sections (left) and the corresponding radial $L_{eff}$ fits (right). %The recorded radial activity distributions (red), averaged over concentric rings around the estimated source location, are fitted by the model (blue) to extract \ldiff. 
Histological types and cell lines are: TNBC 4T1 0.62~g tumor (A,B), PDAC PANC1 0.68~g tumor (C,D), prostate C4-2 0.95~g tumor (E,F), colon HCT116 0.58~g tumor (G,H).}
%\vspace{1 cm}
\label{fig:analysis}
\end{figure}

Four examples of histological sections and the corresponding \ldiff~fits are given in Figure~\ref{fig:analysis}. For each tumor, the left panel shows the PSL distribution across the section, and the right panel shows the ring-averaged \bismuth~specific activity as a function of the radial distance from the estimated source position. The model (blue) is fitted to the data (red) in the region 0.5--2 (or 2.5)~mm as described above. The fit region is selected based on the data distribution to include the exponential fall-off of activity, typically over two orders of magnitude. The radial fit region falls in most cases within 2~mm from the source position. The fit model was always applied in the source midplane (no $z$ offset) as variation of $\pm2~$mm showed no significant impact on the fitted \ldiff~(as discussed also in Ref.~\cite{latticepart1:2023}). Finally, \avgldiff~per tumor was calculated as an average over the histological sections used in the analysis with the uncertainty defined as one standard deviation across the sections. \par 

\subsection{$P_{leak}(Pb)$~measurements}
To determine \leakage, both the total \lead~activity in excised tumors $\Gamma_{Pb}^{tum}$ and the \radium~activity of the extracted source $\Gamma_{Ra}^{src}$ were measured, using a well-type NaI(Tl) detector (Hidex Automatic Gamma Counter). The measurement focused on a region surrounding the 239-keV gamma line of \lead~(intensity 43.6\%), considering (for the source) the contribution of the 241-keV gamma of \radium~(intensity 4.1\%). The activities were corrected to the time of tumor removal, $t_R$. The effective desorption probability $P_{des}^{eff}(Pb)$ was measured for a few selected cases, following a procedure described in Ref~\cite{arazi:2020}, and found to be $0.53\pm0.02$. The measured source activity was within 8\% of the value reported by the manufacturer. The leakage probability was calculated for each tumor as follows, assuming a global value $P_{des}^{eff}(Pb)=0.53$.

\noindent For $t\gg\tau_{Rn}$ the total number of \lead~atoms inside the tumor $N_{Pb}^{tum}(t)$ follows ~\cite{arazi:2020}:
\begin{equation}
    \frac{dN_{Pb}^{tum}}{dt}=P_{des}^{eff}(Pb) \Gamma_{Ra}^{src}(t)-(\lambda_{Pb}+\alpha_{Pb})N_{Pb}^{tum}.
\end{equation}
The source activity is described as:
\begin{equation}
    \Gamma_{Ra}^{src}(t)=\Gamma_{Ra}^{src}(0)e^{-\lambda_{Ra}t}
\end{equation}
and hence the \lead~activity in the tumor as a function of time is given by: 
\begin{equation}
    \Gamma_{Pb}^{tum}(t)=\lambda_{Pb}N_{Pb}^{tum}(t)=
    \frac{\lambda_{Pb}}{\lambda_{Pb}+\alpha_{Pb}-\lambda_{Ra}} P_{des}^{eff}(Pb) \Gamma_{Ra}^{src}(0) \left(e^{-\lambda_{Ra}t}-e^{-(\lambda_{Pb}+\alpha_{Pb})t}\right).
\end{equation}

\noindent For a time of tumor removal ($t_R$) much longer than the \lead~half-life ($t_R \gg \tau_{Pb}$):
\begin{equation} \label{eq:alpha_Pb_from_activities_at_tR}
    \Gamma_{Pb}^{tum}(t_R)=\frac{\lambda_{Pb}}{\lambda_{Pb}+\alpha_{Pb}-\lambda_{Ra}} P_{des}^{eff}(Pb) \Gamma_{Ra}^{src}(t_R).
\end{equation}

\noindent Given $P_{des}^{eff}(Pb)$, by measuring $\Gamma_{Pb}^{tum}(t_R)$ and $\Gamma_{Ra}^{src}(t_R)$, $\alpha_{Pb}$ can be derived from Equation \ref{eq:alpha_Pb_from_activities_at_tR}, and consequently from it the $^{212}$Pb leakage probability as:
\begin{equation}
    P_{leak}(Pb)=\frac{\alpha_{Pb}}{\lambda_{Pb}+\alpha_{Pb}}.
    \label{eq:pleak}
\end{equation}

Now from Equation~\ref{eq:ldiff_lead}, and $\tau_{Pb}^{eff}=\frac{1}{\lambda_{Pb}+\alpha_{Pb}}$, it follows:

\begin{equation*}
    \frac{1}{\lambda_{Pb}+\alpha_{Pb}}=\frac{L_{Pb}^{2}}{D},
\end{equation*}

and from Equation~\ref{eq:pleak}:

\begin{equation}
    P_{leak}(Pb)=1-\frac{\lambda_{Pb}}{\lambda_{Pb}+\alpha_{Pb}}=1-\frac{\lambda_{Pb}L_{Pb}^2}{D_{Pb}}.
    \label{eq:leak_diff}
\end{equation}

In case $L_{Pb}$ is the dominant diffusion length ($L_{eff}\approx L_{Pb}$), this relation should also hold for $L_{eff}$. Therefore, an anti-correlation between $P_{leak}(Pb)$ and $L_{eff}$ can indicate a lead-dominated diffusion scenario. Conversely, if $P_{leak}(Pb)$ is independent of $L_{eff}$, one may conclude that the diffusion is radon-dominated, with $L_{Rn}>L_{Pb}$.

\section{Results}

\subsection{Effective diffusion length in tumors}

The effective diffusion parameter \ldiff~was measured for 113 tumors in total and found to be in the range of 0.2--0.7 mm across different tumor types and sizes, indicating a broad range for different tumors. Only data for prostate cancer cell lines showed a significant increase in \ldiff~with tumor mass ($p\textrm{-value}<0.05$). The measurements of radon diffusion length, \lradon~, for corresponding cell lines~\cite{Heger2024} are consistent with the \ldiff~measurements when performing a one-sided ANOVA test for each histological type (at 0.05 significance). \par

\begin{figure}[h!]
\centering
\includegraphics[width=0.95\textwidth]{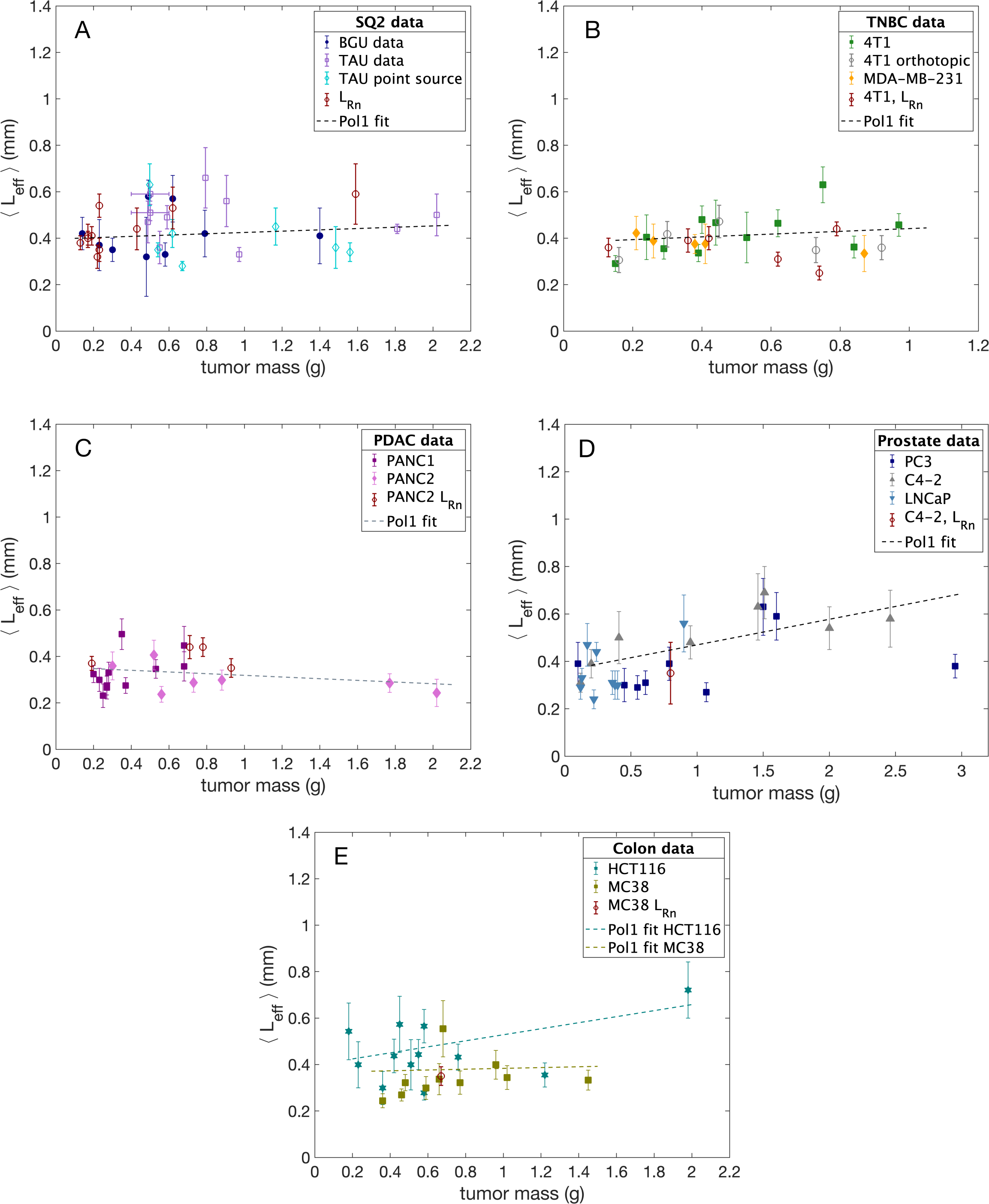}
\caption{Fitted \avgldiff~per tumor as a function of the tumor mass for different histological types and corresponding cell lines: SCC (A), TNBC (B), PDAC (C), prostate cancer (D) and colon cancer (E). SCC results contain additional data sets for comparison (see text for details). Dashed lines represent linear fits and the error bars show the total uncertainty for each tumor measurement.}    
\label{fig:ldiff_mass}
\end{figure}

Figure~\ref{fig:ldiff_mass}A shows \ldiff~for SQ2 tumors as a function of tumor mass (0.2--2~g) for the three protocols described in Section~\ref{sec:methods}: the main protocol (``BGU data'') and the two variations: line sources with varying activities (``TAU data'') and point source measurements (``TAU point source''). \ldiff~values are from 0.3 to 0.7~mm and show no dependence on tumor mass (linear coefficient $0.03 \pm 0.21$ at 95\% CI). The error bars indicate the total uncertainty after combining the statistical uncertainty arising from averaging over histological sections and systematic uncertainty from the iQID calibration correction. The average measured value over all SQ2 tumors is \avgldiff= $0.42 \pm 0.09$~mm. The results for the diffusion length associated with \radon~from Ref. \cite{Heger2024} are added for comparison. One-sided ANOVA indicated no difference between different \ldiff~ samples or \ldiff~and the published $L_{Rn}$ data. The \ldiff~ (for ``BGU data'') and $L_{Rn}$ measurements are also consistent when performing the Mann-Whitney U-test ($p\textrm{-value}=0.7$).

Figure~\ref{fig:ldiff_mass}B shows the results for TNBC for 0.15--0.97~g tumors: 4T1 subcutaneous, 4T1 orthotopic and MDA-MB-231. \ldiff~values range from 0.3--0.6~mm with no apparent mass dependence (linear coefficient $0.06 \pm 0.14$ at 95\% CI). Results for orthotopic and subcutaneous 4T1 tumor models are consistent in the measured data ($p\textrm{-value}=0.3$ for Welch's t-test) and no difference to the human model is observed ($p\textrm{-value}=0.2$ for the same statistic). The average value for 4T1 tumors is \avgldiff= $0.41 \pm 0.09$~mm and for MDA-MB-231, \avgldiff= $0.38 \pm 0.05$~mm. The measured values for the \radon~diffusion length (open red circles) are in agreement with the rest of the data points ($p\textrm{-value}=0.3$ for one-sided ANOVA). 

Results for PDAC tumors are shown in Figure~\ref{fig:ldiff_mass}C for PANC1 and PANC2 models. \ldiff~ranges from 0.23~mm to 0.50~mm, while low values about 0.2~mm are measured even for PANC2 tumors of $\sim$2~g. The average value measured for PANC1 tumors is \avgldiff= $0.33 \pm 0.08$~mm and for PANC2 tumors \avgldiff= $0.30 \pm 0.07$~mm, which are the lowest measured values compared to other histologies. The $L_{\rm{Rn}}$ values (open red circles) are consistent with the rest of the measurements ($p\textrm{-value}=0.1$ for one-sided ANOVA for the three groups).

Data for prostate cancer cell lines---PC3, C4-2 and LNCaP---are shown in Figure~\ref{fig:ldiff_mass}D. The tumors are in the 0.1--3~g mass range and \ldiff~values are between 0.2--0.7~mm with a significant increase with tumor mass (linear coefficient yields $0.11 \pm 0.06$~mm/g at 95\% CI). The average values are $0.36 \pm 0.11$~mm, $0.40 \pm 0.13$~mm and $0.52 \pm 0.13$~mm for LNCaP, PC3 and C4-2, respectively. Only one data point for \radon~measurement is reported and is consistent with the \ldiff~values shown. One-sided ANOVA indicates that the samples are not consistent ($p\textrm{-value}<0.05$) which presumably arises from LNCaP tumors measured at small masses (below 1~g) and C4-2 and PC3 sampled until 3~g. C4-2 tumors yield higher values when compared with PC3 and LNCaP ($p\textrm{-values}$~0.07 and $0.02$, respectively).  \\

Figure~\ref{fig:ldiff_mass}E shows the measured \ldiff~for colon cancer tumors up to 2~g. Human HCT116 model shows systematically larger \ldiff~values compared to its mouse counterpart MC38 ($p\textrm{-value}=0.02$ for Welch's t-test). No significant increase with the tumor mass is observed for both data sets when fitted with linear regression. The average values across all tumors are \avgldiff= $0.45 \pm 0.13$~mm for HCT116 and \avgldiff= $0.34 \pm 0.06$~mm for MC38. One MC38 tumor is shown for $L_{\rm{Rn}}$ measurement (red open circles) and is consistent with other points. The full summary of average values for measured \avgldiff~for each histological type is given in Table~\ref{tab:fuji_avg_results}. The quoted uncertainties are separated in statistical contributions arising from the variations between different tumors and a systematic contribution from the calibration correction. 

%%% commented values for small tumors 
\begin{table}%[ht!]
  \centering
  \begin{tabular}{l|l|c}
  \hline
  \hline
  \textbf{Histological type}	& \textbf{Cell line} & $\langle L_{\rm eff} \rangle \pm \textrm{(stat)} \pm \textrm{(syst)}$  (mm) \\
  \hline
SCC          & SQ2  & $0.42 \pm 0.09 \pm 0.04 $ \\
  \hline

  TNBC	  & MDA-MB-231 &  $0.38 \pm 0.03 \pm 0.03$  \\  
            &	4T1             &  $0.42 \pm 0.09 \pm 0.04$   \\
            &	4T1 orthotopic  &  $0.46 \pm 0.08 \pm 0.04$   \\
  \hline
  PDAC	  & PANC1      &  $0.33 \pm 0.08 \pm 0.03 $ \\
            & PANC2      &  $0.30 \pm 0.06 \pm 0.03 $  \\

  \hline

  Prostate cancer	& PC3 & $0.40 \pm 0.13 \pm 0.04 $ \\
            & C4-2 & $0.52 \pm 0.13 \pm 0.05 $  \\
            & LNCaP & $0.36 \pm 0.10 \pm 0.03 $  \\
  \hline
  Colon cancer	& HCT116 & $0.45 \pm 0.13 \pm 0.04 $ \\
        & MC38   & $0.34 \pm 0.05 \pm 0.03 $ \\
  \hline\hline
  \end{tabular}
  \caption{Average~\avgldiff~ across all measured tumors for each cell line model studied in the experiment. Statistical uncertainty is one SD across all tumors and the systematic uncertainty arises from the calibration correction described in Appendix~\ref{app:calibration}. }
\label{tab:fuji_avg_results}
\end{table}

\subsection{Leakage measurement}
\begin{figure}[h!]
\centering
\includegraphics[width=0.95\textwidth]{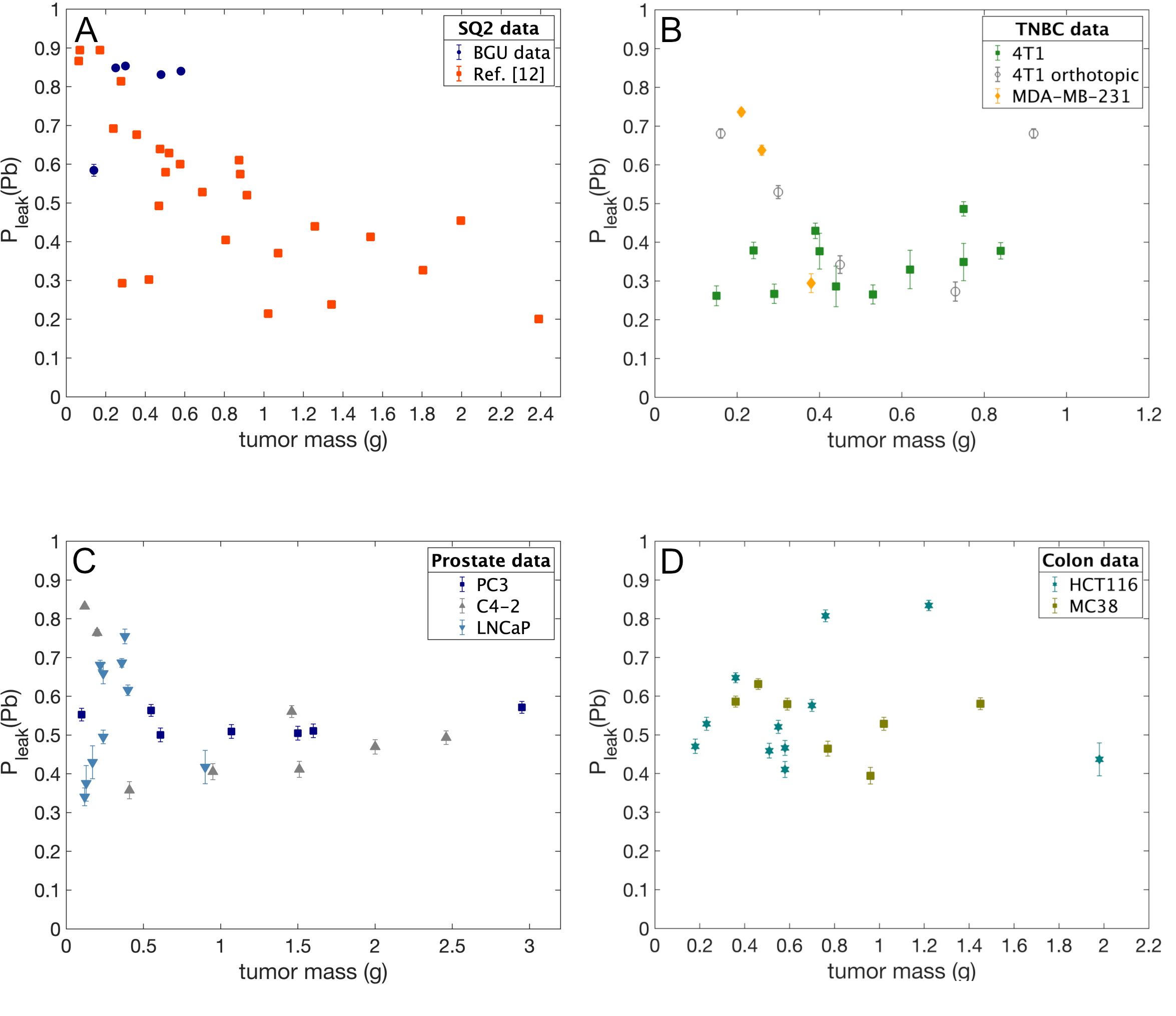}   
\caption{Measured \leakage~per tumor as a function of the tumor mass for different histological types and corresponding cell lines: SCC (A), TNBC (B), prostate cancer (C) and colon cancer cell lines (D). SCC results from Ref.~\cite{arazi:2007} are added to panel A for completion.} 
\label{fig:leakage_mass}
\end{figure}

% explain here how you did it and what is the result.
\leakage~was measured for 77 tumors originating from 10 cell lines (part of the dataset used for \ldiff) and found to span a wide range from 0.1 to 0.9, with varying dependence on tumor mass. Figure~\ref{fig:leakage_mass}~shows \leakage~vs. tumor mass for different cell lines. In panel A, similarly measured SQ2 data (26 tumors) from Ref. \cite{arazi:2007} show a significant decrease in \leakage~with tumor mass. TNBC data in panel B suggest that for subcutaneous 4T1 tumors \leakage~is mass-independent, while for orthotopic 4T1 and MDA-MB-231 \leakage~generally decreases with tumor mass. Only 3 data points were measured for pancreatic tumors in the range from 0.5 to 0.8. For prostate cancer, PC3 tumors show no mass dependence with $P_{leak}(Pb)\sim0.5$ on average, while C4-2 and LNCaP tumors display an apparent decrease of leakage with mass below $\sim0.5$~g. Lastly, for colon cancer tumors (D) there is no observed dependence of leakage on tumor mass, with $P_{leak}(Pb)\sim0.5$. Correlations between the measured \leakage~and \avgldiff~per tumor were inspected for each cell line. Equation~\ref{eq:leak_diff}, which holds for \ldiff~instead of \llead~ if the diffusion is \lead-dominated in secular equilibrium, predicts a negative correlation between the variables. However, a statistically significant negative correlation was observed only in MC38 and C4-2 tumors when inspected with two-dimensional linear regression and Pearson's correlation coefficient, as described in Appendix \ref{app:correlation}.

%\clearpage

\section{Implication for the dose around a single source}

The observed root-mean-square spread in the measured values of \ldiff~for a given tumor type ranges from 8\% to 33\%, with the largest variations recorded for prostate PC3 tumors. This spread can result (at least partially) from local variations in necrotic domains (which are more prevalent in larger tumors) and vascular density, which is typically higher in the periphery of the tumor. Since the Alpha-DaRT source in these experiments is not much shorter than the tumor size, different sections along the source axis can present strong variations in \ldiff. Incorporating such variations into a realistic dosimetry model is far from trivial and requires knowledge of the size distribution of tissue domains with different diffusion and leakage characteristics. As a first step towards estimating the effect of the observed variations, we adopt a simple and pragmatic approach in which the tissue around the source is uniform (such that the simple DL model equations hold with constant coefficients) but the values of the diffusion lengths and leakage probability take a range of values consistent with the observed spread.

\begin{figure}[!]
\centering
\includegraphics[width=0.95\textwidth]{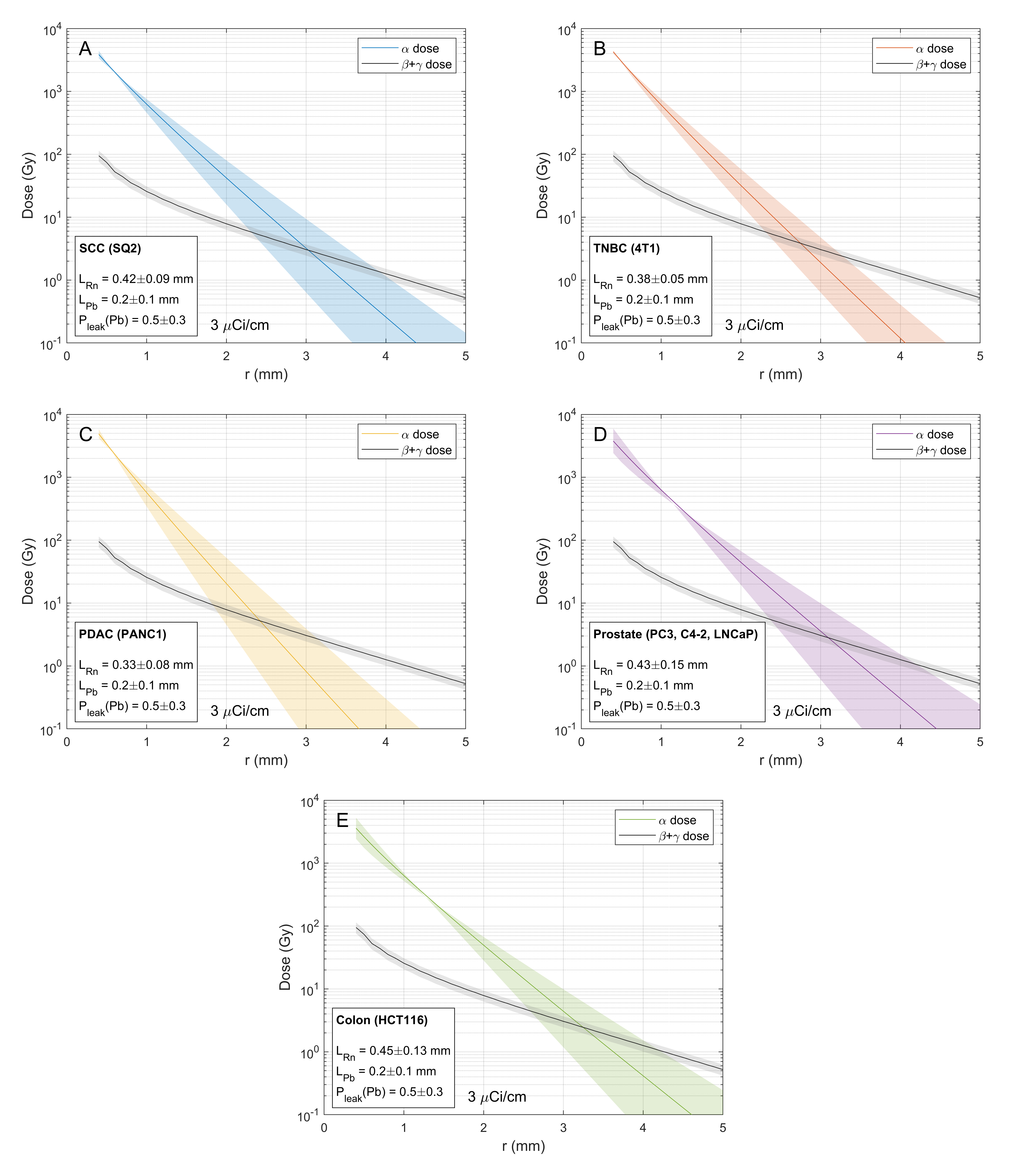}   
\caption{Alpha and low-LET radial dose profiles for the studied tumor types, based on the observed variations in $L_{eff}$ and \leakage: SCC (A), TNBC (B), PDAC (C), prostate (D) and colon (E). Diffusion is assumed to be radon-dominated with \ldiff~representing $L_{Rn}$. The source, in all cases, carries 3~$\mu$Ci/cm $^{224}$Ra and is 0.7 mm in diameter. The desorption probabilities are $P_{des}(Rn)=0.45$ and $P_{des}^{eff}(Pb)=0.55$. The low-LET dose is calculated by approximating the source to a line and assuming that all electron and photon emissions originate from it with no diffusion. Lines are for the nominal values of \lradon, \llead~and \leakage, with the bands corresponding to variations of all parameters within the ranges shown in the legends.} 
\label{fig:dose_profile_bands}
\end{figure}

Figure \ref{fig:dose_profile_bands} shows how the observed variations in \ldiff~and \leakage~affect the calculated dose profile in the midplane of the source under this simplistic scenario (all sources are assumed to be 0.7~mm in diameter, with 3~$\mu$Ci/cm $^{224}$Ra, $P_{des}(Rn)=0.45$, and $P_{des}^{eff}(Pb)=0.55$). The calculation was done assuming that \ldiff~represents $L_{Rn}$, with a spread of values corresponding to $\pm$ one standard deviation. The diffusion is assumed to be radon-dominated, with $L_{Pb}=0.1-0.3$~mm and $P_{leak}(Pb)=0.2-0.8$. (The nominal values are taken as $L_{Pb}=0.2$~mm and $P_{leak}(Pb)=0.5$.) Each panel shows a band corresponding to the radial profile of the alpha dose (in color) and low-LET dose in grey. The latter was calculated by approximating the source to a line, and assuming that all electron and photon emissions originate from the source (i.e., using the ``line-source/no-diffusion'' approximation discussed in our previous publication\cite{epstein2024}, and shown to be accurate to $\sim10-15\%$ when compared to a full-diffusion simulation). Since we neglect diffusion when calculating the low-LET dose, the low-LET dose profile is the same for all tumor types. Variations in the model parameters lead to variations in the radial distance at which the asymptotic alpha dose reaches a certain reference value (e.g., 10~Gy). For SCC tumors, the 10-Gy alpha dose level is reached at $r=2.6\pm0.4$~mm, for TNBC at $r=2.4\pm0.3$~mm, for PDAC at $r=2.2\pm0.4$~mm, for prostate tumors at $r=2.6\pm0.4$~mm, and for colon tumors at $r=2.7\pm0.3$~mm. Variations across tumor types are thus fairly small. A 1-cm-long Alpha-DaRT source carrying 3~$\mu$Ci $^{224}$Ra per cm length creates a cylindrical region with a nominal diameter of $4.4-5.4$~mm in which the asymptotic alpha dose is higher than 10~Gy.

\section{Discussion}
\label{sec:discussion}
In this study, we report on \textit{in vivo} measurements of the effective diffusion length \ldiff~and the \lead~leakage probability~\leakage. \ldiff~was measured in 113 mice across 10 cell lines and was found to lie in the range 0.2–0.7~mm for tumor masses between 0.1 and 3~g. Measured \leakage~values in a total of 77 tumors are between 10–90\% with a smaller spread for larger tumors around $\sim50\%$. While these data are taken from mice-borne tumors (of both murine and human origin), this is at present the only way to estimate the DL model parameters towards treatment planning in clinical trials, since no external imaging or internal dose measurements techniques can provide the spatial resolution required to extract the sub-mm effective diffusion length. 

The consistency between most measured values for \ldiff~and the previously published results for {$L_{Rn}$} suggests that in those cases, the diffusion is radon-dominated rather than lead-dominated, or that the two diffusion lengths are of similar magnitude. Based solely on the DL model, for the measured range of \ldiff~(0.2--0.7~mm) and the previously reported $L_{Rn}$ values between 0.3--0.5~mm, the estimated $L_{Pb}$ could lie in the 0.2--0.5~mm range (for the fit region of 0.5--2~mm from the source). It is important to note that in the lead-dominated diffusion case, one would expect to find a negative correlation between the measured \leakage~and \ldiff.  Since such correlations are apparently observed in some measured cell lines (MC38 and C4-2 tumors), some contribution of \lead~diffusion cannot be excluded. An increase in \ldiff~as a function of tumor mass observed in some cell lines may be understood by additional effects of tissue necrosis and considerable \lead~contribution. A plausible explanation for the observed decrease in $^{212}$Pb leakage with tumor mass is that for large tumors, a source placed roughly in the tumor center lies farther away from vascular-rich regions in the tumor periphery, and hence the average clearance time of $^{212}$Pb is longer. The role of vasculature could further be investigated using histopathology and imaging techniques.  \par

To probe in more detail the interplay between the diffusion lengths, more \textit{in vivo} measurements are required: experiments expanding the histology and tumor mass coverage of \lradon~measurements as well as tackling the vascular contribution further away from the source. More orthotopic models are being envisioned, such as for colon and brain cancer.

Based on the presented results of parameter measurements, it appears that a choice of 4-mm spacing (preferably in a hexagonal source configuration) offers a reasonable starting point for ongoing and future clinical trials. For SCC, treatments employing sources carrying 2~$\mu$Ci/cm $^{224}$Ra placed at a nominal spacing of 5~mm resulted in a complete response (CR) rate of 78.6\%. Reducing the nominal spacing to 4~mm and increasing the $^{224}$Ra activity to 3~$\mu$Ci/cm further improved the CR rate to $\sim90\%$, with a two-year local recurrence-free survival of 77\%. Similar results may be expected in other tumor types, especially if their corresponding EBRT prescription dose is not much higher than that of SCC and if they are easily accessible from the outside (e.g., breast cancer). However, treating hard-to-reach tumors such as in the pancreas or colon metastases in the lungs or liver, may require increased source activities to compensate for the difficulty in precise source placement. A similar increase in activity may also be needed for prostate tumors, where---although hexagonal source placement is feasible---the perscription dose may be much higher. 

\section{Conclusion}
\label{sec:conclusion}

This study presented results from preclinical \textit{in vivo} experiments with Alpha-DaRT for the purpose of providing a reasonably informed starting point for treatment planning in ongoing and future clinical trials on various tumor types. The measurements indicate that the diffusion spread around the Alpha-DaRT source is radon-dominated. With the estimated values of DL model parameters extracted from these studies, a feasible treatment plan can be established that provides therapeutic dose coverage across the hexagonal \dart source lattice. Such plans should consider the effect of tumor-specific parameters and the required biological effective dose when setting source spacing and activities. The presented data is a valuable guide for initial planning in Alpha-DaRT clinical protocols with an increasingly growing number of studies being planned for the near future.

%%%%%%%%%%%%%%%%%%%%%%%%%%% APPENDIX %%%%%%%%%%%%%%%%%%%%%%%%%
\clearpage
\begin{appendices}

\section{Detailed description of materials and methods}
\label{app:methods}
\subsection{Cell culture and tumor inoculation}

\paragraph{Cell culture}
Cells were grown in 5\% CO${_2}$ in a humidified incubator at 37$^{\circ}$~C. All media were supplemented with 10\% fetal bovine serum (FBS), penicillin (100 U/ml), and streptomycin (100 $\mu$g/ml) (all obtained from Gibco, Thermo Fisher Scientific, MA, USA). The SQ2 (squamous cell carcinoma, SCC), MDA-MB-231 and 4T1 (both triple negative breast cancer, TNBC), PANC2 (pancreatic duct adenocarcinoma, PDAC), 
and MC38 (colon carcinoma) cell lines were provided by Prof. Yona Keisari (Sackler School of Medicine, Tel-Aviv University, Israel). In total, 9 SQ2 tumors, 16 4T1, 5 MDA-MB-231, 8 PANC2 and 10 MC38 tumors were grown for this study. The three prostate carcinoma cell lines - PC3 (9 tumors), C4-2 (8 tumors) and LNCaP (9 tumors) - were provided by Dr. Benesova Martina (DKFZ-Heidelberg, German Cancer Research Center Foundation, Heidelberg, Germany). The PANC1, PDAC cell line, was provided by Prof. Moshe Oren (Moross Integrated Cancer Center, Weizmann Institute of Science, Rehovot, Israel; 11 tumors). The colon carcinoma cell line HCT116 (in total 12 tumors) was provided by Prof. Bert Vogelstein (Sidney Kimmel Comprehensive Cancer Center, Johns Hopkins University, MD, USA). All cells were grown in Dulbecco's Modified Eagle Medium (DMEM), except for 4T1 and the three prostate carcinoma cell lines, which were grown in Roswell Park Memorial Institute Medium (RPMI), and HCT116 cells, which were grown in McCoy's 5A Medium. The number of instilled cells varied from 5$\cdot10^{5}$ in the case of 4T1 and SQ2 tumors to 5$\cdot10^{6}$ in the case of PANC1 and the prostate carcinoma cells. In addition to these samples, autoradiography data from a previous measurement campaign (denoted as ``TAU data"), including 17 SQ2 tumors, were added for comparison. 

\begin{figure}
 \begin{minipage}[b]{\linewidth}
        %\centering
        \includegraphics[width=\textwidth]{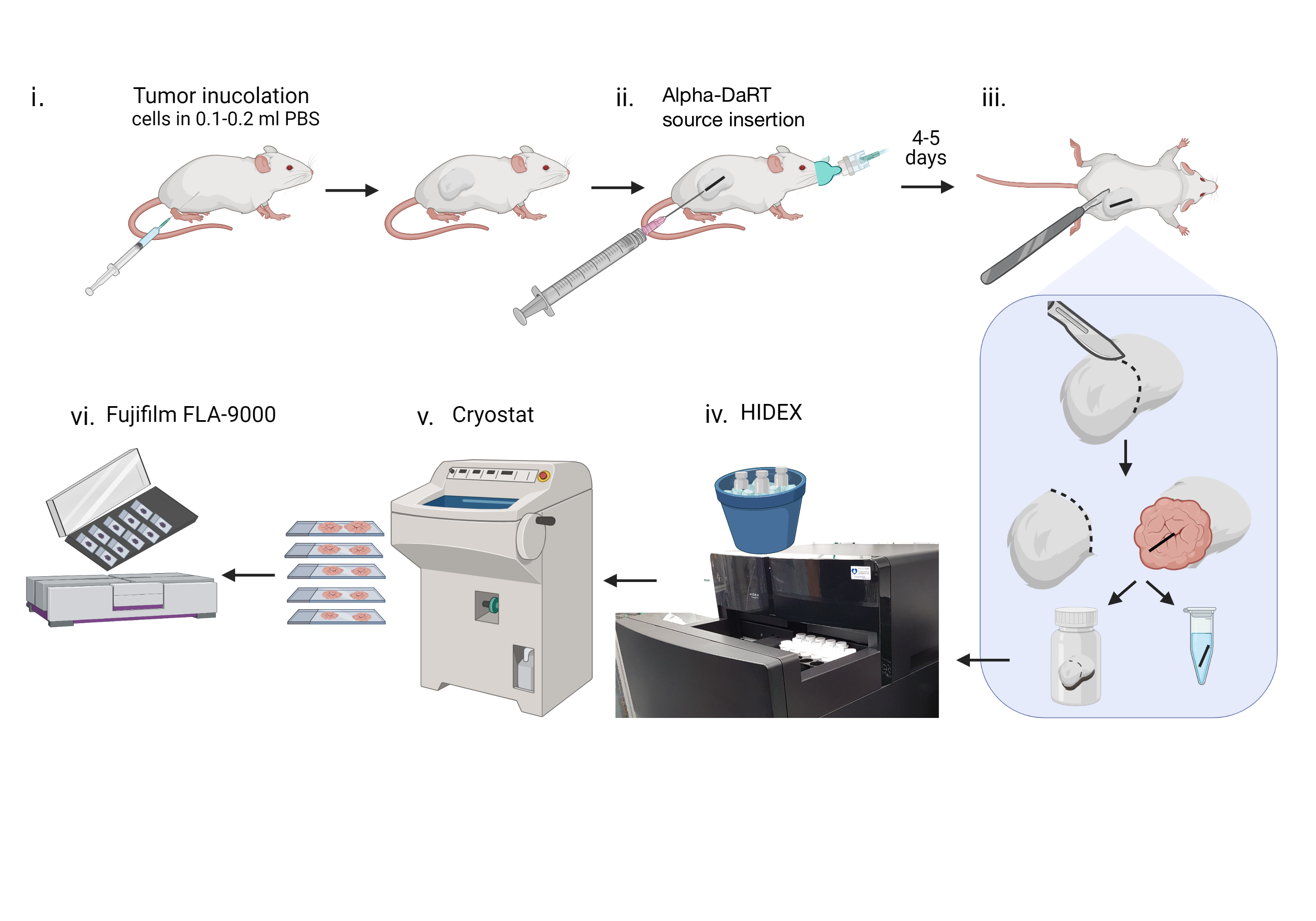}
    \end{minipage}
 \caption{Experimental sequence: (i) tumor inoculation – cells were injected subcutaneously in 0.1-0.2 ml PBS; after a period of 8-50 days of tumor development (depending on the cell line growth kinetics) (ii) the tumors were treated with a single Alpha-DaRT source insertion into the middle of the tumor while mice were anesthetized. Four to five days later (iii) the tumors were removed and cut in the middle, and the source was extracted and placed in 1.5 ml tube full of water. The tumors were weighed, placed in 20~ml scintillation bottles on dry ice and taken to the Hidex gamma counter for activity measurement (iv). Following source removal, both halves of the tumor were subjected to histological frozen-sectioning by LEICA CM 1520 cryostat (v). The 10~$\mu$m-thick sections were then placed on positively charged glass slides, with 250-300~$\mu$m intervals between each section, creating a series of sequential sections (typically, between 5 to 15 slides per tumor). Finally, after fixation, slides were taken to the autoradiography system, Fujifilm FLA-9000 (vi).}
 \label{fig:methods}
\end{figure}

\paragraph{Tumor inoculation}
The study was approved by the Ben-Gurion University Institutional Animal Care and Use Committee, and was conducted according to the Israeli Animal Welfare Act following the guidelines of the Guide for Care and Use of Laboratory Animals (National Research Council, 1996) [permit no. IL-78-12-2018(E) and IL-44-08-2021 (E)]. BALB/c mice (6-12 weeks old) were obtained from Envigo (Jerusalem, Israel). NOD scid gamma (NSG) mice (6-12 weeks old) were obtained from the breeding colony of Ben-Gurion University, Beer-Sheva, Israel. C57BL/6 mice were bred ‘in-house’ at the animal facility of Ben-Gurion University. Mice were inoculated subcutaneously with cells in 100-200$~\mu$l Dulbecco's phosphate-buffered saline (DPBS) (Gibco, 14190144, Thermo Fisher Scientific, MA, USA) into the low lateral side of the back. Depending on the target tumor mass, the inoculation-to-treatment time varied between 8-99 days. Figure~\ref{fig:methods} shows the experimental sequence starting with tumor inoculation in (i).

\paragraph{Alpha-DaRT source insertion}
Alpha-DaRT sources made out of 316LVM stainless steel loaded with 3~$\mu$Ci \isotope[224]{Ra} (0.7 mm in diameter and 6.5 mm long), were placed near the tip of an 18-gauge needle attached to a 2.5 ml syringe (Picindolor, Rome, Italy) and inserted into the tumor by a plunger placed internally along the syringe axis (Figure~\ref{fig:methods} (ii)). Source insertion was done under anesthesia with isoflurane. Source location was verified using a Geiger counter (RAM GENE-1, Rotem industries, Israel) after insertion. 

\paragraph{Frozen section preparation}
Four to five days post Alpha-DaRT treatment, the tumors were excised (as a whole). Each tumor was cut to two halves, in the estimated location of the source center, perpendicular to the source’s insertion axis as shown in Figure~\ref{fig:methods} (iii). The source was then pulled out using surgical tweezers and was placed in a 1.5 ml microcentrifuge tube filled with 1 ml of water. The tumor was placed for 1 hour in dry ice at -80$^{\circ}$C and then moved to a 20 ml scintillation bottle on dry-ice and taken for total \lead~activity measurement in a Hidex Automatic Gamma Counter  (Figure~\ref{fig:methods} (iv)).  

Following the activity measurement, both halves of the tumor were subjected to histological frozen sectioning by a LEICA CM 1520 cryostat (Buffalo Grove, IL, USA), Figure~\ref{fig:methods} (v). The 10-$\mu$m-thick sections were then placed on positively charged glass slides, with 250-300 $\mu$m intervals between each section, creating a series of sequential samples (between 5 to 15 per tumor). Following the sectioning, slides were fixed with 4\% paraformaldehyde (sc-281692, Santa Cruz Biotechnology Inc., Dallas, Texas, USA) for 10 minutes and rinsed twice with PBS for 10 minutes each time. 
Immediately after the fixation step, slides were taken to the autoradiography system (Figure~\ref{fig:methods} (vi)).

\subsection{Phosphor imaging calibration}
\label{app:calibration}

The conversion from gray-level (GL) to photo-stimulated luminescence (PSL) level is given by the image reader and is of the form $I_{\rm{PSL}}=\textrm{exp}(a \, I_{\rm{GL}}+b)$, where $a, b$ depend on the specific scanner resolution and $I$ is the image matrix. The recorded PSL level intensity patterns were converted to activity using \lead~calibration samples measured separately. The samples were prepared by the following procedure. A small cylindrical plastic container (28.5 mm diameter and 3 mm height) is placed inside a low-pressure chamber ($<$0.1~mbar) with a \thorium~source at the bottom of the container and an aluminium sheet glued to its top. Over several days, \radium~ions recoiling from the \thorium~source are implanted in the aluminium sheet. Then, the \radium~implanted container top is removed and placed on top of a second container, holding a plastic sheet at the bottom, which is covered by an adhesive tape with a circular 5~mm-diameter hole in the middle. In this configuration, the recoiling daughters of \radium~are allowed to hit the exposed circular part of the plastic sheet over a prescribed duration (typically several hours), creating a circular patch of \lead~with a typical activity in the range 5-200~Bq. Once this stage is finished, the adhesive tape is removed, leaving only the circular exposed plastic as the active calibration sample, which is then taken to the Hidex gamma counter to determine its \lead~ activity. Finally, the source, with its back side glued to a glass slide, is taken to the Fuji system to be measured along with the tumor sections.  A total of four calibration measurements were carried out, yielding a linear relation between PSL and \lead~activity. 

\begin{figure}[ht]
    \centering
    \includegraphics[width=0.7\textwidth]{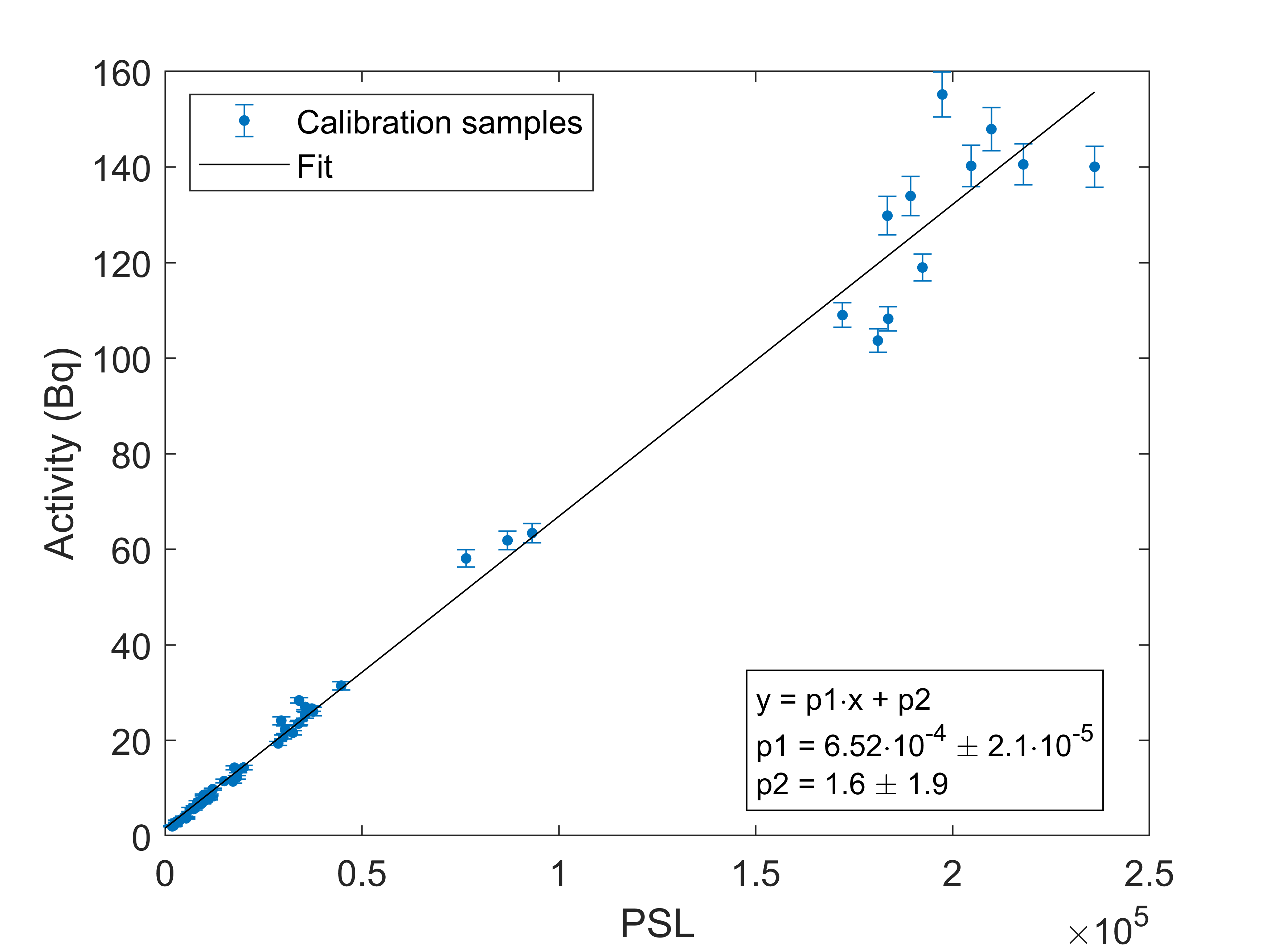}
    \caption{Phosphor imager PSL-to-activity calibration curve.}
\label{figure:PI_calibration}
\end{figure}

\begin{figure}[ht]
    \centering
    \includegraphics[width=1\textwidth]{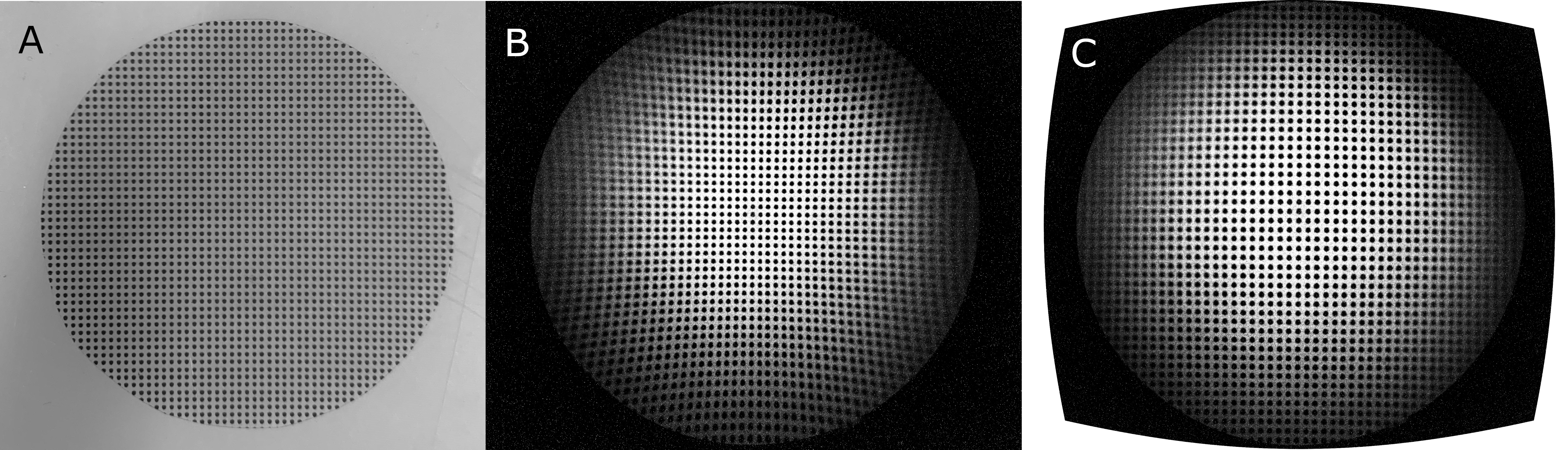}
    \caption{(A) the grid pattern used for the distortion correction. (B) the raw image of the grid pattern acquired using the iQID system. (C) the corrected grid image.}
\label{figure:grid_correction}
\end{figure}

To account for possible machine-dependent differences in image production, raw PSL images recorded by the phosphor-imaging system were cross-calibrated with the iQID camera~\cite{miller2014} using the exact same tumor sections.
The iQID operates as follows: the samples are placed face-down on a scintillating material - in this case, a sheet of ZnS:Ag (Eljen technologies, EJ-440-100). Each alpha particle impinging on the scintillating sheet releases photons. These pass through an image intensifier, and, following a series of lenses, imaged by a CCD sensor. Each flash of light is referred to as a ``cluster''. The cluster images are constantly sent to the PC operating the iQID, where each cluster's center is identified. The XY coordinates, as well as the timestamp of each cluster, are stored in a list file. Since this analysis occurs on an event-by-event basis, there is no convolution with each cluster's point spread function (PSF) in the final images and, thus, no need for de-convolution when analyzing the iQID images. The resulting image is then corrected for geometrical aberrations using the \texttt{lensdistort()} MATLAB script \cite{lensdistort}. The correction parameter is obtained by imaging a printed grid pattern with known distances between the grid points, such as the one shown in Figure \ref{figure:grid_correction}. After applying the correction, the average distance between dots in the image acquired with the iQID is accurate to 0.08\%.

\begin{figure}
    \centering
    \includegraphics[width=\textwidth]{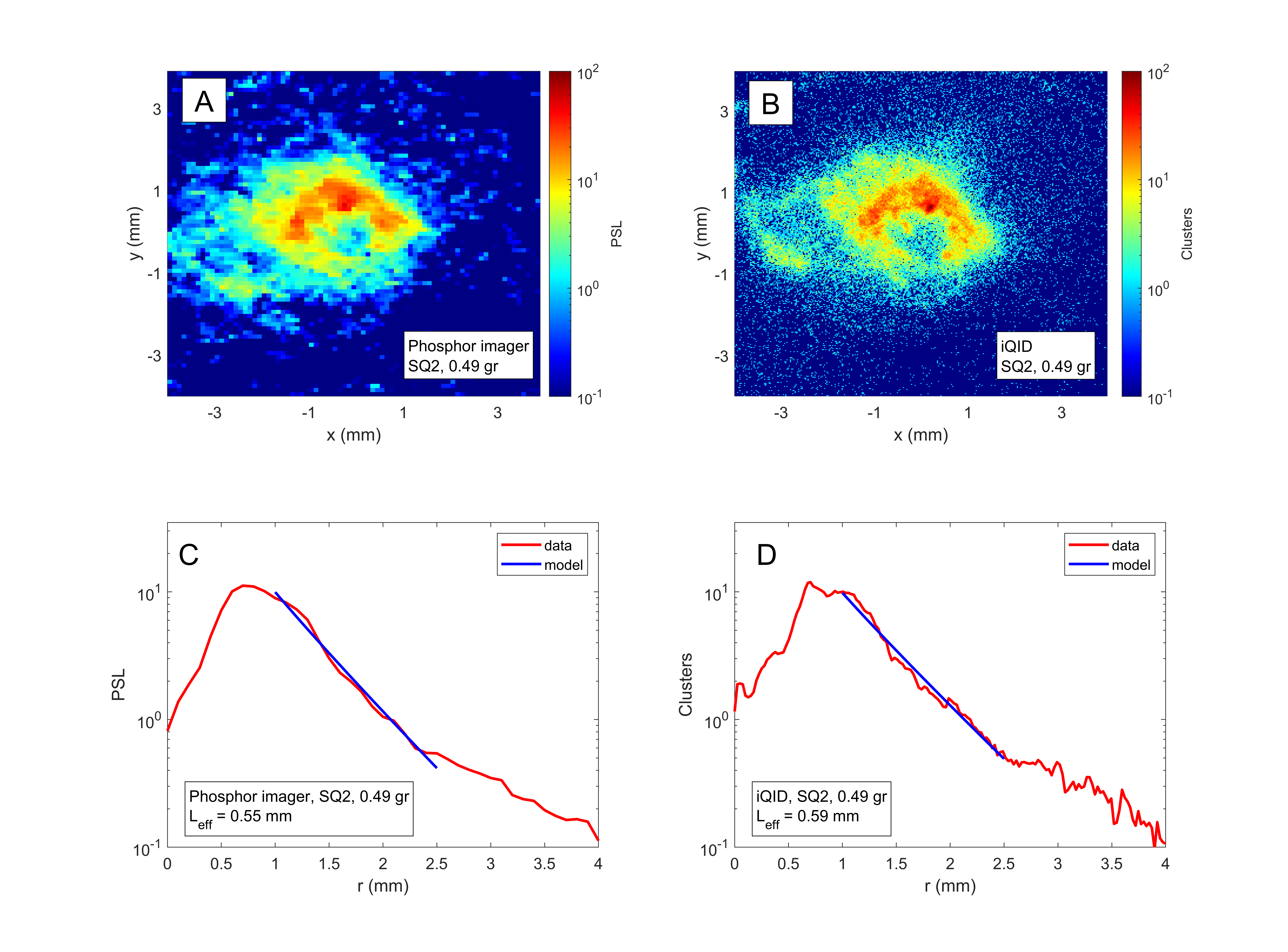}
    \caption{Comparison between the iQID and phosphor-imaging systems: (A) PSL image of a section measured in the phosphor imaging system. (B) cluster image of the same section measured in the iQID system. (C) PSL profiles (measured and fitted) and extracted diffusion length from the phosphor-imaging measurements. (D) the same, for the iQID measurement.}
 \label{fig:iqid_fuji_comp}
\end{figure}

Figure \ref{fig:iqid_fuji_comp} shows an example of one section, from a 0.49 gr, SQ2 mice-borne tumor, measured in both the phosphor imaging system (A) and the iQID system (B). Additionally, it shows the extracted activity profiles and diffusion lengths from the respective measurement system (C)-(D). As seen in this example, the \ldiff~value extracted from the raw PSL image from the phosphor imaging system is lower than the iQID value. This is true for 90\% of the measured sections. The reason for this could be some sharpening of the image being applied internally by the phosphor imaging system. In total, 77 histological sections from six tumors were imaged for comparison between the systems. Averaging over all sections, the resulting ratio is: $L_{iQID}/L_{phosphor~imager} = 1.09 \pm 0.11$. This correction factor is thus applied to all measured phosphor-imaged samples accounting for its uncertainty (i.e., the \ldiff~value extracted from the raw PSL image is multiplied by this factor).

\color{black}

\subsection{Correlation between diffusion and leakage}
\label{app:correlation}

\begin{figure}
\centering
\includegraphics[width=0.8\textwidth]{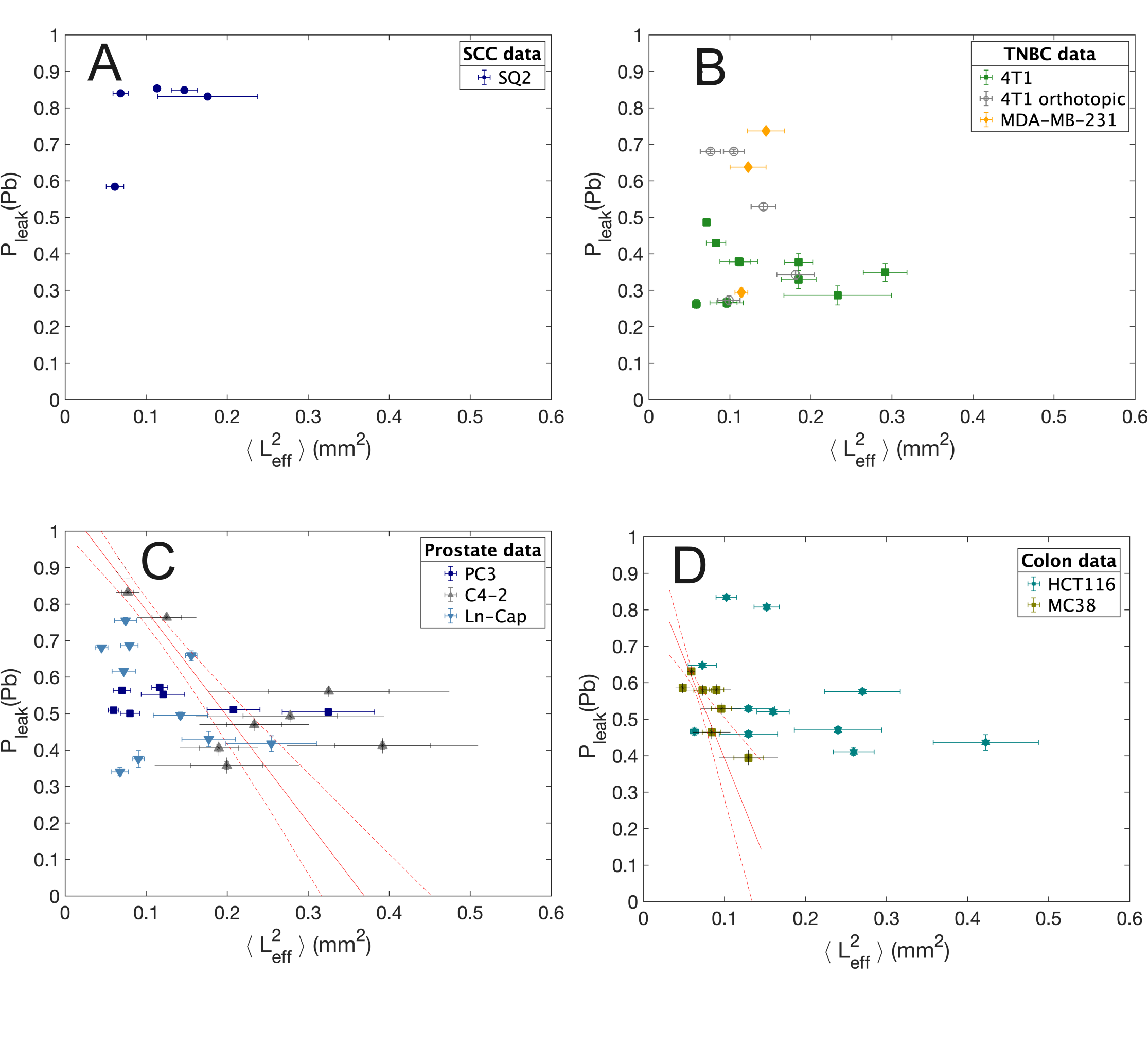}
\caption{Calculated \leakage~per tumor as a function of the \avgldiff~for different histological types and corresponding cell lines: SCC (A), TNBC (B), prostate cancer (C) and colon cancer (D). Dashed lines represent the linear fit to measured points for C4-2 and MC38 cell lines together with 95\% CI. The error bars represent statistical uncertainties of $\pm1\sigma$.} \label{fig:correlations}
\end{figure}

The possible correlation between the measured \leakage~and $\langle L_{eff}^2 \rangle$ was inspected for all cell lines by using a linear fit, minimizing the $\chi^{2}$ calculated using uncertainties in both variables~\cite{browaeys2022}. In addition, Pearson's correlation coefficient $\rho$ and the corrected Pearson's correlation coefficient $\rho'$ were calculated for each cell line data. The correction to the Pearson coefficient was calculated by considering the uncorrelated additive uncertainties for \leakage~and \avgldiff. Statistically significant negative correlation was obtained for C4-2 ($p_{1}=-3.0 \pm 0.7$ $\rm{mm}^{-1}$, $\rho=-0.63$, $\rho'=-0.69$ with $p$-value 0.10) and MC38 ($p_{1}=-5.4 \pm 2.8$ $\rm{mm}^{-1}$, $\rho=-0.83$, $\rho'=-0.89$ with $p$-value 0.02). Figure~\ref{fig:correlations} shows the measured correlations for all analyzed histologies with relevant fits indicated for C4-2 and MC38 data.

\end{appendices}

\section*{References}
\bibliography{bibliography}
\bibliographystyle{./medphy.bst}

\end{document}